\documentclass[aps,prl,twocolumn,showpacs,superscriptaddress,groupedaddress,nofootinbib,floatfix]{revtex4-1}

\usepackage{graphicx,psfrag}
\usepackage{mathrsfs}
\usepackage{amsmath,amsfonts,amssymb}
\usepackage{multirow}
\usepackage{comment}
\usepackage{ulem}
\usepackage{hyperref}

\newcommand{\be}{\begin{equation}}
\newcommand{\ee}{\end{equation}}
\newcommand{\bea}{\begin{eqnarray}}
\newcommand{\eea}{\end{eqnarray}}
\newcommand{\bel}{\begin{align}}
\newcommand{\eel}{\end{align}}

\def\ergsec{{\rm erg\,s^{-1}}}

\def\Msun{{\rm M_{\odot}}}
\def\GMc2{{\rm G M_{\odot} c^{-2}}}

\def\O{\mathcal{O}}
\def\Mpr{M_\text{pc}}
\def\Cpr{c_\text{pc}}
\def\egw{e_\text{GW}}
\def\Egw{E_\text{GW}}

\usepackage{color}

\definecolor{cyan}{rgb}{0,0.9,0.9}
\definecolor{orange}{rgb}{0.9,0.5,0}
\definecolor{magenta}{rgb}{1,0,1}
\definecolor{purple}{rgb}{0.8,0.4,0.8}
\definecolor{gray}{rgb}{0.8242,0.8242,0.8242}

\begin{document}

\title{Gravitational-wave luminosity of binary neutron stars mergers}

\author{Francesco \surname{Zappa}$^{1}$}
\author{Sebastiano \surname{Bernuzzi}$^{1,2}$}
\author{David \surname{Radice}$^{3,4}$}
\author{Albino \surname{Perego}$^{2,1,5}$}
\author{Tim \surname{Dietrich}$^6$}

\affiliation{${}^1$ Dipartimento di Scienze Matematiche Fisiche ed Informatiche, Universit\'a di Parma, I-43124 Parma, Italia}
\affiliation{${}^2$ Istituto Nazionale di Fisica Nucleare, Sezione Milano Bicocca, gruppo collegato di Parma, I-43124 Parma, Italia}
\affiliation{${}^3$ Institute for Advanced Study, 1 Einstein Drive, Princeton, NJ 08540, USA}
\affiliation{${}^4$ Department of Astrophysical Sciences, Princeton University, 4 Ivy Lane, Princeton, NJ 08544, USA}
\affiliation{${}^5$ Dipartimento di Fisica, Universit\`{a} degli Studi di Milano Bicocca, Piazza della Scienza 3, 20126 Milano, Italia}
\affiliation{${}^6$ Max Planck Institute for Gravitational Physics (Albert Einstein Institute), Am M\"uhlenberg 1, Potsdam-Golm, 14476, Germany} 

\date{\today}

\begin{abstract}
  We study the gravitational-wave peak luminosity and radiated energy of
  quasicircular neutron star mergers using a large sample of numerical relativity 
  simulations with different binary parameters and input physics. 
  The peak luminosity for all the binaries can be described in terms of the mass 
  ratio and of the leading-order post-Newtonian tidal parameter solely.
  The mergers resulting in a prompt collapse to black
  hole have largest peak luminosities. However, the largest amount of energy per unit mass is radiated by mergers
  that produce a hypermassive neutron star or a massive neutron star remnant.
  We quantify the gravitational-wave luminosity of binary
  neutron star merger events, and set upper limits on the radiated
  energy and the remnant angular momentum from these events.  
  We find that there is an empirical universal relation connecting the total
  gravitational radiation and the angular momentum of the remnant.
  Our results constrain the final spin of the remnant black-hole
  and also indicate that stable neutron star remnant forms with
  super-Keplerian angular momentum.
\end{abstract}


\maketitle

Gravitational waves (GWs) from a likely binary neutron star (BNS)
inspiral have been observed for the first time on August 17th 2017 
during the
second observational run of Advanced LIGO and Virgo \cite{TheLIGOScientific:2017qsa}. The
observation sets a lower bound to the total radiated energy, 
$\Egw>0.025\Msun {\rm c}^2$, by considering only a portion of the GW signal
corresponding to the inspiral dynamics. 
The largest GW energy, however, is expected to be radiated during the merger and the 
subsequent postmerger phases~\cite{Bernuzzi:2014kca,Bernuzzi:2015rla}. The
only way to theoretically quantify the emitted GW energy is to perform
numerical relativity (NR) simulations. NR-based models can be then
evaluated on the intrinsic parameters of the binary esimated from the
observations to obtain the emitted energy. 
In this work we study the GW peak
luminosity and GW energy emitted by quasicircular binary neutron star
mergers using one of the largest set of NR simulations currently
available \cite{Bernuzzi:2014kca,Bernuzzi:2014owa,Dietrich:2015iva,Bernuzzi:2015rla,Dietrich:2017feu,Dietrich:2017aum,
Radice:2016rys,Dietrich:2016hky,Dietrich:2016lyp, Radice:2017zta, Radice:2017lry, 
RadiceInPrep}.

Compact binary mergers are the most powerful events
in the Universe in terms of GW energy. The binary
black hole (BBH) mergers observed so far emitted about $1-3~\Msun~{\rm c}^2$ 
with peak luminosities reaching $200~\Msun {\rm c}^2~{\rm s}^{-1}$ (about
$\sim3-4\times10^{56}~\ergsec$)
\cite{Abbott:2016blz,Abbott:2016nmj,Abbott:2017vtc}.
The largest luminosity is reached for an equal-masses and aligned spins
configuration, with both holes spinning at maximum rate. Physically, spin-orbit interactions during
the dynamics enhance the emission for the spin aligned
configurations. Fits to the BBH luminosity and radiated energy as function of
mass ratio and spins have been developed in a number of NR-based
works,~e.g.~\cite{Baker:2008mj,Keitel:2016krm,Jimenez-Forteza:2016oae,Healy:2016lce}.  
By constrast, the total radiated GW energy of BNS has been quantified only
for particular cases,~e.g.~\cite{Baiotti:2008ra,Bernuzzi:2015rla,Dietrich:2016hky,Dietrich:2016lyp},
and quantitative models for predicting the properties of the merger remnant are missing.

\begin{figure*}[t]
  \centering 
  \includegraphics[width=\textwidth]{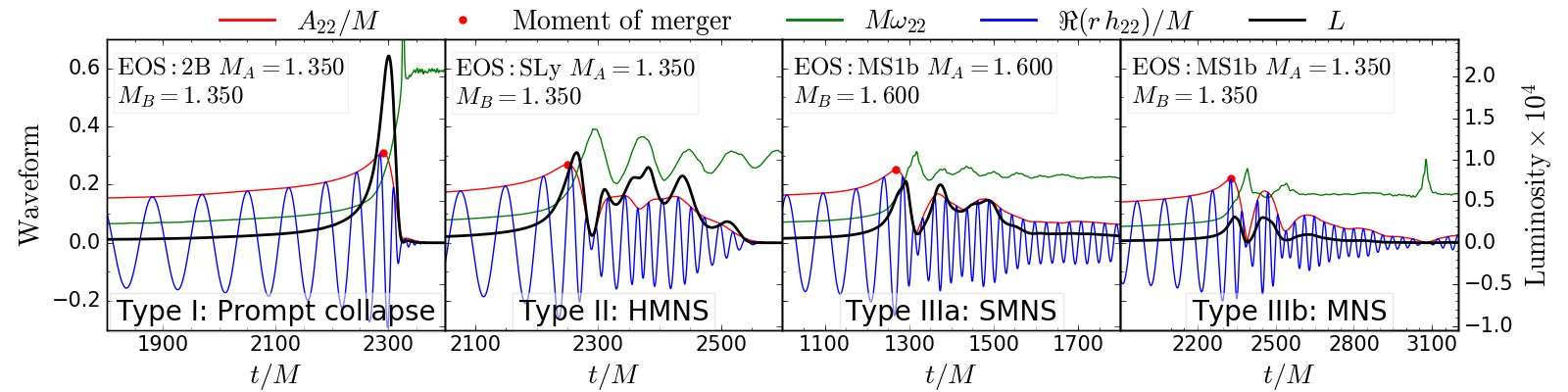}
  \caption{Merger waveforms and GW luminosity for the three types of
    mergers. From left to right the GW correspond to a merger ending
    in: a prompt collapse to black hole (Type I), a hypermassive
    neutron star (Type II), a supramassive neutron star (Type IIIa)
    and a massive stable neutron star (Type IIIb). Note the double y-axis.}
  \label{fig:waves}
\end{figure*}

We consider 100 different BNS simulations that include variation of the 
gravitational binary mass $M=M_A+M_B\in[2.4, 3.4]\Msun$, the mass ratio $q=M_A/M_B\in[1, 2.06]$, and a sample of
8 equations of state (EOSs) comprising 4 finite-temperature
microphysical EOS models. Spin interactions in about 30 BNS are
simulated consistently in general relativity following
\cite{Tichy:2011gw}. Spins are either aligned or
antialigned to the orbital angular momentum, and of varying magnitude up to 
$|\mathbf{S}|/M^2\approx0.15$.
A microphysical treatment of neutrino cooling is included in 37
simulations, following the method presented in~\cite{Radice:2016dwd}.
Four simulations also included an effective treatment of
turbulent angular momentum transport that may arise from small scale
magnetohydrodynamical instabilities in the merger remnant
\cite{Radice:2017zta}.
Most of the BNSs are simulated at multiple grid resolutions for a total
of more than 200 datasets, that
guarantee control on numerical artifacts. Simulations are performed with the BAM~\cite{Brugmann:2008zz,*Thierfelder:2011yi} and THC
codes~\cite{radice:2013hxh, *radice:2013xpa}.
Full details on the data are given elsewhere~\cite{DietrichInPrep}. 
The GW energy $E_{\rm GW}$ and the binary's angular momentum $J$ are
calculated from our simulations from the GW multipolar waveform, as
described in~\cite{Damour:2011fu,Bernuzzi:2014kca}. 
We work with the mass and symmetric mass-ratio, $\nu=M_A M_B/M^2$, rescaled
quantities, $\egw=\Egw/(M \nu)$ and $j=J/(M^2\nu)$.
The luminosity peak is computed as $L_{\rm peak} = \max_t
\left\{ d\Egw(t)/dt \right\}$. Note that, differently from BBH, the BNS
luminosity does depend on the binary mass due to tidal interactions
during the dynamics (see below). 
The conversion factor from geometric units $G=c=M_\odot=1$ used here to CGS units is the
Planck luminosity
\be
L_P = \frac{c^5}{G} \approx 3.63 \times 10^{59}\ \ergsec \ ;
\ee
the typical order of magnitude of $L_{\rm peak}$ for
compact binary mergers is $10^{-3}-10^{-4}L_P$.

The BNS merger dynamics is crucially determined by tidal
interactions~\cite{Bernuzzi:2014kca,Bernuzzi:2015rla}. 
Ref.~\cite{Bernuzzi:2014kca} has shown that $\egw$, $j$ and many other key quantities at the
moment of merger~\footnote{The moment of merger is formally defined as the time of the waveform amplitude's peak, that corresponds to the end of the chirp signal.} 
can be fully characterized by the sum 
\be\label{eq:kap2T}
\kappa^T_2=\kappa_2^{\rm A}+\kappa_2^{\rm B} \ ,
\ee
of the gravitoelectric quadrupolar tidal polarizability 
coefficients~\cite{Damour:2009wj}
\be\label{eq:kappaA}
\kappa_2^{\rm A} = 2\frac{X_B}{X_A}\left(\frac{X_A}{C_A}\right)^5 k^A_2 \ .
\ee
Above, $k^{\rm A}_2$ is the quadrupolar Love number describing the static
quadrupolar deformation of body $A$ in the gravitoelectric field of
the companion, 
$C_A$ is the compactness, 
and $X_A=M_A/M$. The coefficient  
$\kappa^T_2$ parametrizes at leading-order the tidal 
interactions in the general-relativistic
2-body Hamiltonian, waveform's phase and amplitude~\cite{Damour:2012yf}.
Larger energy emissions correspond to smaller values of $\kappa_2^T$,
that, in turn, gets smaller values for larger masses, more compact NSs and
softer EOS. In what follows we
show that a similar characterization holds also for the peak
luminosity. 

The possible outcomes of a BNS merger are a prompt collapse to black hole
(Type~I), a hypermassive NS (HMNS, Type~II), a supramassive NS (SMNS,
Type~IIIa), or a stable NS (MNS,
Type~IIIb)~\cite{Baumgarte:1999cq,Hotokezaka:2011dh,Bauswein:2013jpa}. 
We find that the GW peak luminosity is reached during merger and the
subsequent dynamical phase and it strongly depends on the merger
type. For Type~I mergers the luminosity peak just follows the moment
of merger, similarly to the BBH case. Type~II mergers have 
multiple peaks of comparable luminosity on a time scale of
$\O(100M)$ (few~ms). The peaks following
the moment of merger are related to the HMNS emission and can be
of comparable or 
stronger magnitude. Type~III mergers are qualitatively similar to Type~II,
but the peak luminosities are lower. Four representative simulations are presented in Fig.~\ref{fig:waves}.

\begin{figure}[t]
  \centering 
  \includegraphics[width=0.49\textwidth]{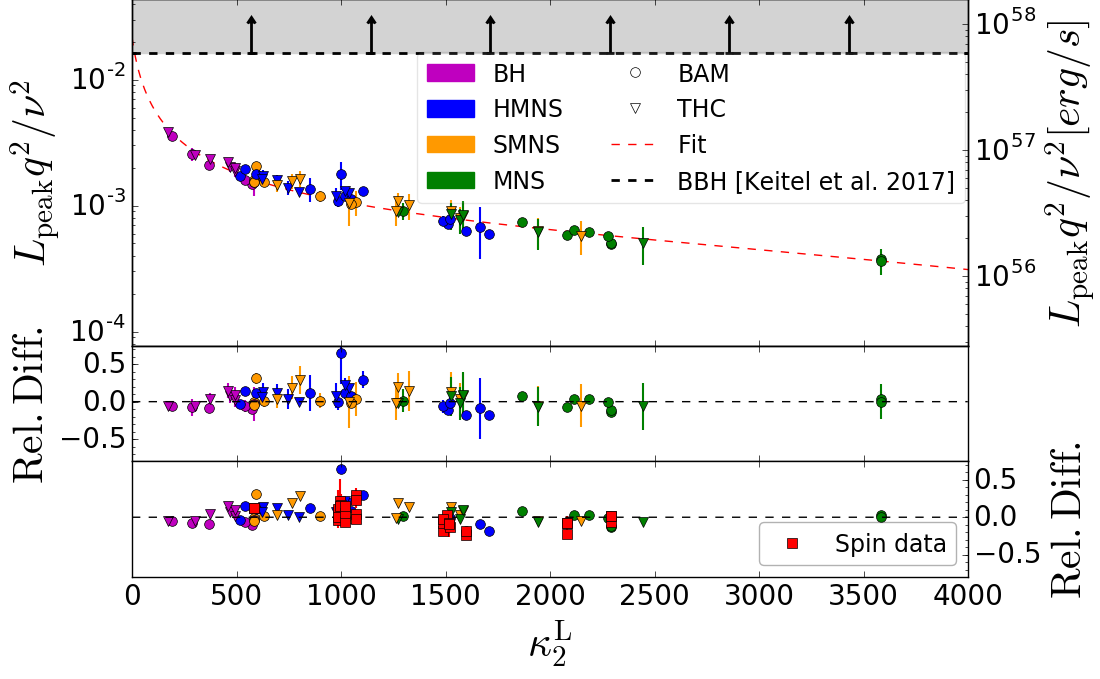}
  \caption{GW luminosity peak as a function of the tidal parameter
    $\kappa^{\rm L}_2$.
    The errorbars are calculated from simulations performed
    at different resolutions. Second panel: fit's residuals with errors. 
    Third panel: residuals of spin data compared to the fit.}
  \label{fig:Lfit2}
\end{figure}

The BNS peak luminosity can be characterized by a simple function of
the tidal polarizability coefficients, Eq.~\eqref{eq:kappaA}. In the
post-Newtonian (PN) description of the inspiral 
dynamics, tidal effects contribute to the luminosity with a leading
order 5PN term  $\delta L_{\rm Tidal}  = \frac{32}{5} \nu^2 x^{10} \kappa_2^{\rm L}$
\cite{Hinderer:2009ca}, where $x = (\pi M f_{\rm GW} )^{2/3}$ is the PN
expansion parameter, $f_{\rm GW}$ is the GW frequency and 
\be\label{eq:kappaL}
\kappa_2^{\rm L} = 
2\left[
\frac{3 - 2 X_A}{X_B} \kappa^A_2 + 
(A \leftrightarrow B)
\right]  \ .
\ee
The perturbative parameter $\kappa_2^{\rm L}$ captures the
strong-field dynamics behaviour for
$L_\text{peak}$ as shown in Fig.~\ref{fig:Lfit2}. Our irrotational BNS
sample can be fit by~\footnote{Similar results are obtained also
  using $\kappa^T_2$ since $X_A\sim X_B\sim1/2$.}
\be
\label{L_fit1}
L_\text{peak}(\nu,\kappa^\text{L}_2) \approx L_0 \frac{\nu^2}{q^2(\nu)} 
\frac{( 1 + n_1 \kappa^\text{L}_2 + n_2 ({\kappa^\text{L}_2})^2 )}
     {( 1 + d_1 \kappa_2^\text{L})} \ ,
\ee
with $L_0 = 2.178\times{10^{-2}},\ n_1 = 5.2(4)\times{10^{-4}},\ n_2
= -9.3(6)\times{10^{-8}},\ d_1 = 2.7(7)\times{10^{-2}}$ and a coefficient
of determination $R^2=0.943$. The maximal residuals are of the order of 
$30\%$ (with one outlier at $\sim60\%$). Note that the prediction using BBH 
fits would overestimate $L_\text{peak}$ of, at least, a factor 4. Our fit also captures the
spinning BNS data. For the spin magnitudes considered here, 
the spin-orbit contribution to $L_\text{peak}$ is
within the fit and numerical uncertainties.
As an example of application, a fiducial equal-mass BNS with $M=2.8\Msun$ and
$\kappa^A_2=\kappa^B_2\sim92$ ($\kappa_2^\text{L}\sim1472$)
has $L_\text{peak}\sim 8.168\times10^{-4}$
($\sim1.852\times10^{55}\ \ergsec$). The application of
Eq.~\eqref{L_fit1} to GW170817 is also straightforward and just 
requires to evaluate the posteriors for the likely distribution of the 
mass ratio the tidal parameters.

The $L_\text{peak}$ analysis also highlights that the threshold between
Type~I and Type~II mergers is approximately controlled by the value of
$\kappa^\text{L}_2$ (or $\kappa^\text{T}_2$).
Prompt collapse happens above a mass threshold $M>\Mpr=\Cpr M^{\rm TOV}_{\rm max}$, 
where $M^{\rm TOV}_{\rm max}$ is the maximum gravitational mass 
of a nonrotating NS and $1.3 \lesssim \Cpr \lesssim 1.6$ is a constant
that depends only weakly on the binary's mass-ratio. 
Both $\Cpr$ and $M^{\rm TOV}_{\rm max}$ depend on the
EOS~\cite{Hotokezaka:2012ze,Hotokezaka:2013iia,Bauswein:2013jpa}.
For a given EOS, the prompt collapse threshold translates into limiting values
of $\kappa_{2\ {\rm pc}}^\text{T}$ (or $\kappa_{2\ {\rm pc}}^\text{L}$), that can be computed by
considering all the possible pairs of NS such that $M_A+M_B=\Mpr$
(with $1.1~\Msun<M_\text{A}<M^{\rm TOV}_{\rm max}$). For our set of 8 EOS 
we find that
Type~I mergers are characterized by $\kappa^\text{T}_{2\ {\rm pc}}\sim80$ 
($\kappa^\text{L}_{2\ {\rm pc}}\sim600$) 
where the value can vary of about
$\delta\kappa^\text{T}_{2\ {\rm pc}}\lesssim40$
($\delta\kappa^\text{L}_{2\ {\rm pc}}\lesssim200$). 
Such predictions are verified by our NR sample, although no common
threshold can be found for all the considered EOS.

\begin{figure}[t]
  \centering 
  \includegraphics[width=.5\textwidth]{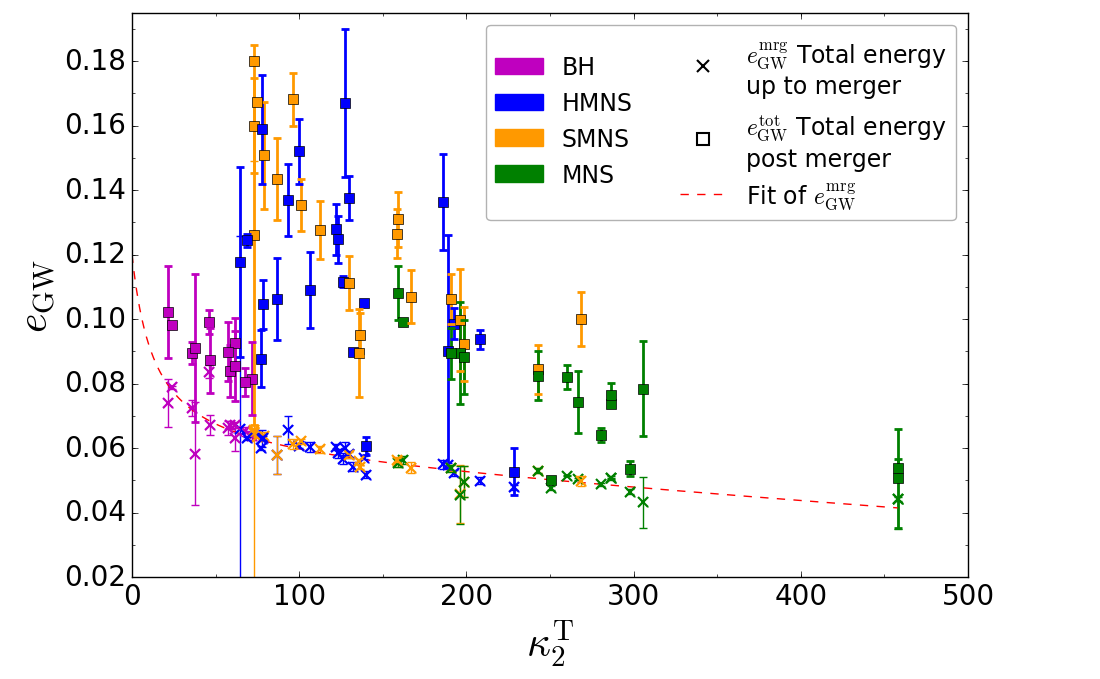}
  \caption{Reduced GW energy at merger ($\times$) and total ($\square$) as a function
    of the tidal parameter $\kappa^\text{T}_2$.} 
  \label{fig:egw}
\end{figure}

The most luminous BNS do not correspond, in general, to the BNS that radiate the
largest amount of energy. That is yet another difference with respect
to BBH. The largest GW energies {\it per unit mass} are radiated by Type~II
mergers over typical timescales of few tens of milliseconds after the
moment of merger~\cite{Bernuzzi:2015opx}. The remnant HMNS undergoing
gravitational collapse is a very efficient emitter of
GWs; about two times the energy emitted during the inspiral and merger
can be emitted during the postmerger phase. Figure~\ref{fig:egw} shows
the total energy, $\egw^\text{tot}$, and the energy irradiated up to the moment of
merger, $\egw^\text{mrg}$, as a function of
$\kappa^\text{T}_2$ for our irrotational BNS sample. 
While $\egw^\text{mrg}$ tightly
correlates with $\kappa^\text{T}_2$, the total energy has a
more complex behaviour. Our results set an upper bound of 
$\egw^{\rm tot}\lesssim0.18$ obtained for $100\lesssim
\kappa^\text{T}_2\lesssim200$, e.g.~for the fiducial $M=2.8\Msun$ BNS
discussed above. Hence, if two different BNSs with 
$M\sim2.8\Msun$ and $\nu\sim~1/4$ are Type~I and Type~II
respectively~\footnote{For example if the NS matter is softer/stiffer
  in one case.}, then the former might be louder and the latter 
might emit more energy. Given similar total masses, Type~II mergers 
can be louder of a factor $\egw^{\rm tot}({\rm Type~II})/\egw^{\rm
  tot}({\rm Type~I})\sim1.8$ with respect to Type~I, and of a factor 3
with respect to Type~III. 
However, not all Type~II are louder then Type~I. Sufficiently large
individual NS masses in Type~I mergers can rescale $\egw^{\rm tot}$ to
larger absolute energies than those of Type~II.
The largest GW energy that a BNS can emit can be inferred from our
dataset, we find
\be
\Egw^{\rm tot}\lesssim 0.126\frac{M}{2.8}~\Msun {\rm c^2} \ .
\ee
Our results implies that current LIGO-Virgo GW searches at 
kiloHertz-frequencies are insensitive to the 
postmerger signal (Cf. Fig.~1 of~\cite{Abbott:2017eaw}).  

\begin{figure}[t]
  \centering 
  \includegraphics[width=.5\textwidth]{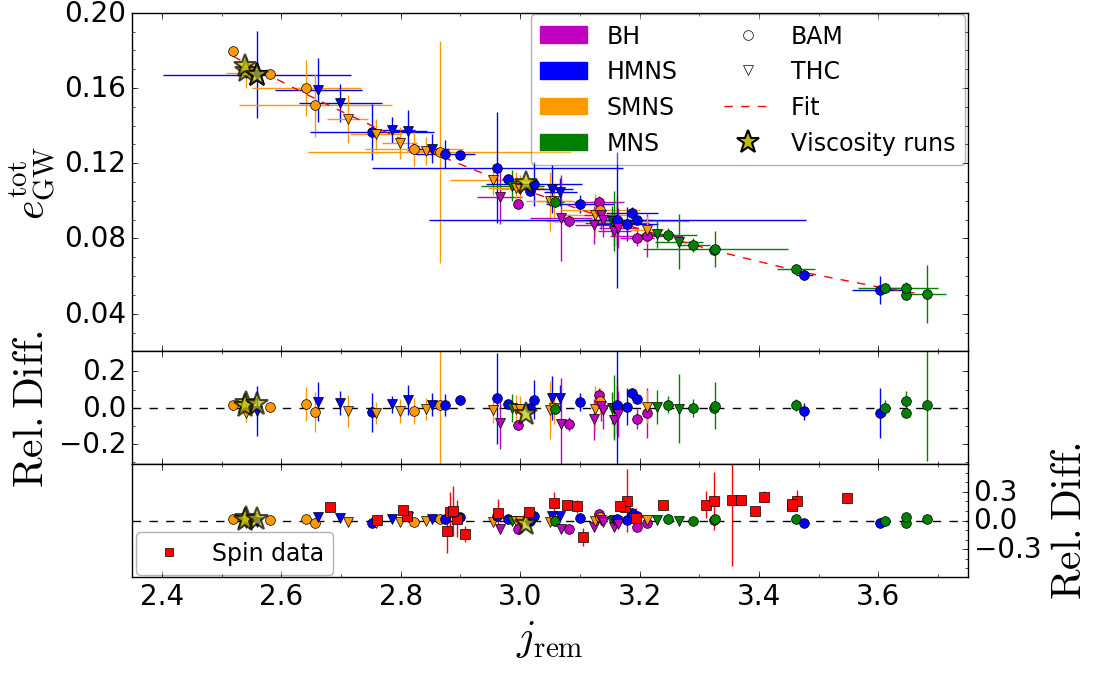}
  \caption{Reduced total GW energy vs. final angular momentum of the remnant.
  Simulations including magnetically-driven viscosity
  effects~\cite{Radice:2017zta} are marked with stars, the empty star
  refers to a control run with zero viscosity. Lower panels: residuals of 
  the fit and the residuals of spin data, with errors.}
  \label{fig:energy_momang}
\end{figure}

Finally, we show that the total radiated energy uniquely determines the
angular momentum of the merger remnant, cf.~Fig.~\ref{fig:energy_momang}. All
the BNS remnants are characterized by values that lay on a given $\egw^{\rm
  tot}(j_{\rm rem})$ curve. That happens rather independendently from the
binary's intrisinc parameters but also from the particular physics simulated in the
postmerger. Notably, the simulations employing viscosity and
neutrino cooling (marked with stars in the plot) lay on the same curve
of simulations employing a purely hydrodynamical prescription for the
matter~\cite{Radice:2017zta}. 
This fact suggests that the emission of gravitational radiation is the dominant
mechanisms determining the dynamics on the dynamical timescales after
merger, $T_\text{dyn}\sim20$~ms.
The irrotational NR data are well described by the relation,  
\be
\label{eq:ejf}
\egw^{\rm tot} \approx c_2 j^2_\text{rem} - c_1 
j_\text{rem} + c_0  \ , 
\ee
where $c_0 = 0.9(4), c_1 = -0.4(3), c_2 = 0.05(3)$, with fit residuals below $20\%$. Spinning data increase fit
residuals to $30\%$.

For Type~I and II mergers, the final spin of the remnant black hole
can be estimated from the angular momentum of the remnant system (BH or HMNS + disk) 
at the end of the initial, GW dominated phase. Thus, Eq.~\eqref{eq:ejf} could be used 
to estimate the final BH spin from the measurement of the energy
radiated by the binary in GWs, which might be possible with third-generation GW observatories.
The value of $J_\text{rem}/M^2$ provides
an upper limit for the remnant BH dimensionless spin, we predict 
$0.6\lesssim J_\text{rem}/M^2\lesssim0.9$ for 
moderatly spinning BNS.
Type~I mergers produce the smallest disks ($\sim 10^{-3}\Msun$), carrying a 
negligible amount of angular momentum \cite{Shibata:2006nm, Rezzolla:2010fd, Shibata:2017xdx, Radice:2017lry}.
Thus, the remnant and final BH angular momenta coincide and 
$0.75 \lesssim \left( J/M^2 \right)_{\rm BH,Type~I} \lesssim0.8$, where 
the fastest spinning black holes are associated with larger values of $\kappa_2^{\rm L}$.
For Type~II mergers, we estimate that a disk of baryon mass 
$M_{b,{\rm disk}}\sim0.1~\Msun$ contains 10-15\% of $J_{\rm rem}$.
Viscosity-driven disk ejecta can carry away a large fraction of this momentum over the disk lifetime
while we evaluate that $\Delta( J/M^2 )_{\rm BH,Type~II} \lesssim 0.03$ by accretion.
For the final BH dimensionless spin we predict $0.6\lesssim \left( J/M^2 \right)_{\rm BH,Type~II}\lesssim0.85$,
where the slowest spinning BHs are produced by light, symmetric BNSs.

The dimensionless angular momentum at the end of the initial, GW
dominated, phase of the postmerger evolution for Type~III binaries is in
the range $0.62 \lesssim J_\text{rem}/M^2 \lesssim 0.82$.
We compare $J_\text{rem}$ for each Type-III binary to that of sequences of uniformly rotating NSs having the same rest-mass. We find that $J_\text{rem}$ exceeds, in most cases significantly, the Keplerian limit. Type~III remnants are thus {\it super-Keplerian}. This suggests that 
the subsequent viscous evolution is likely to be accompanied by massive
outflows \cite{Fujibayashi:2017puw, RadiceInPrep}.

\paragraph*{Acknowledgments}
  We thank A.~Nagar for comments. 
  SB acknowledges support by the EU H2020 under ERC Starting Grant, 
  no.~BinGraSp-714626. 
  DR acknowledges support from a Frank and Peggy Taplin Membership at the
  Institute for Advanced Study and the
  Max-Planck/Princeton Center (MPPC) for Plasma Physics (NSF PHY-1523261).
  Computations were performed on the supercomputer SuperMUC at the LRZ
  (Munich) under the project number pr48pu, on the supercomputers
  Bridges, Comet, and Stampede (NSF XSEDE allocation TG-PHY160025), on
  NSF/NCSA Blue Waters 
  (NSF PRAC ACI-1440083), and on Marconi (PRACE proposal 2016153522 and
  ISCRA-B project number HP10B2PL6K).


\bibliography{references,local}

\begin{thebibliography}{43}%
\makeatletter
\providecommand \@ifxundefined [1]{%
 \@ifx{#1\undefined}
}%
\providecommand \@ifnum [1]{%
 \ifnum #1\expandafter \@firstoftwo
 \else \expandafter \@secondoftwo
 \fi
}%
\providecommand \@ifx [1]{%
 \ifx #1\expandafter \@firstoftwo
 \else \expandafter \@secondoftwo
 \fi
}%
\providecommand \natexlab [1]{#1}%
\providecommand \enquote  [1]{``#1''}%
\providecommand \bibnamefont  [1]{#1}%
\providecommand \bibfnamefont [1]{#1}%
\providecommand \citenamefont [1]{#1}%
\providecommand \href@noop [0]{\@secondoftwo}%
\providecommand \href [0]{\begingroup \@sanitize@url \@href}%
\providecommand \@href[1]{\@@startlink{#1}\@@href}%
\providecommand \@@href[1]{\endgroup#1\@@endlink}%
\providecommand \@sanitize@url [0]{\catcode `\\12\catcode `\$12\catcode
  `\&12\catcode `\#12\catcode `\^12\catcode `\_12\catcode `\%12\relax}%
\providecommand \@@startlink[1]{}%
\providecommand \@@endlink[0]{}%
\providecommand \url  [0]{\begingroup\@sanitize@url \@url }%
\providecommand \@url [1]{\endgroup\@href {#1}{\urlprefix }}%
\providecommand \urlprefix  [0]{URL }%
\providecommand \Eprint [0]{\href }%
\providecommand \doibase [0]{http://dx.doi.org/}%
\providecommand \selectlanguage [0]{\@gobble}%
\providecommand \bibinfo  [0]{\@secondoftwo}%
\providecommand \bibfield  [0]{\@secondoftwo}%
\providecommand \translation [1]{[#1]}%
\providecommand \BibitemOpen [0]{}%
\providecommand \bibitemStop [0]{}%
\providecommand \bibitemNoStop [0]{.\EOS\space}%
\providecommand \EOS [0]{\spacefactor3000\relax}%
\providecommand \BibitemShut  [1]{\csname bibitem#1\endcsname}%
\let\auto@bib@innerbib\@empty
\bibitem [{\citenamefont {Abbott}\ \emph
  {et~al.}(2017{\natexlab{a}})\citenamefont {Abbott} \emph
  {et~al.}}]{TheLIGOScientific:2017qsa}%
  \BibitemOpen
  \bibfield  {author} {\bibinfo {author} {\bibfnamefont {B.~P.}\ \bibnamefont
  {Abbott}} \emph {et~al.} (\bibinfo {collaboration} {Virgo, LIGO
  Scientific}),\ }\href {\doibase 10.1103/PhysRevLett.119.161101} {\bibfield
  {journal} {\bibinfo  {journal} {Phys. Rev. Lett.}\ }\textbf {\bibinfo
  {volume} {119}},\ \bibinfo {pages} {161101} (\bibinfo {year}
  {2017}{\natexlab{a}})},\ \Eprint {http://arxiv.org/abs/1710.05832}
  {arXiv:1710.05832 [gr-qc]} \BibitemShut {NoStop}%
\bibitem [{\citenamefont {Bernuzzi}\ \emph {et~al.}(2014)\citenamefont
  {Bernuzzi}, \citenamefont {Nagar}, \citenamefont {Balmelli}, \citenamefont
  {Dietrich},\ and\ \citenamefont {Ujevic}}]{Bernuzzi:2014kca}%
  \BibitemOpen
  \bibfield  {author} {\bibinfo {author} {\bibfnamefont {S.}~\bibnamefont
  {Bernuzzi}}, \bibinfo {author} {\bibfnamefont {A.}~\bibnamefont {Nagar}},
  \bibinfo {author} {\bibfnamefont {S.}~\bibnamefont {Balmelli}}, \bibinfo
  {author} {\bibfnamefont {T.}~\bibnamefont {Dietrich}}, \ and\ \bibinfo
  {author} {\bibfnamefont {M.}~\bibnamefont {Ujevic}},\ }\href {\doibase
  10.1103/PhysRevLett.112.201101} {\bibfield  {journal} {\bibinfo  {journal}
  {Phys.Rev.Lett.}\ }\textbf {\bibinfo {volume} {112}},\ \bibinfo {pages}
  {201101} (\bibinfo {year} {2014})},\ \Eprint {http://arxiv.org/abs/1402.6244}
  {arXiv:1402.6244 [gr-qc]} \BibitemShut {NoStop}%
\bibitem [{\citenamefont {Bernuzzi}\ \emph
  {et~al.}(2015{\natexlab{a}})\citenamefont {Bernuzzi}, \citenamefont
  {Dietrich},\ and\ \citenamefont {Nagar}}]{Bernuzzi:2015rla}%
  \BibitemOpen
  \bibfield  {author} {\bibinfo {author} {\bibfnamefont {S.}~\bibnamefont
  {Bernuzzi}}, \bibinfo {author} {\bibfnamefont {T.}~\bibnamefont {Dietrich}},
  \ and\ \bibinfo {author} {\bibfnamefont {A.}~\bibnamefont {Nagar}},\ }\href
  {\doibase 10.1103/PhysRevLett.115.091101} {\bibfield  {journal} {\bibinfo
  {journal} {Phys. Rev. Lett.}\ }\textbf {\bibinfo {volume} {115}},\ \bibinfo
  {pages} {091101} (\bibinfo {year} {2015}{\natexlab{a}})},\ \Eprint
  {http://arxiv.org/abs/1504.01764} {arXiv:1504.01764 [gr-qc]} \BibitemShut
  {NoStop}%
\bibitem [{\citenamefont {Bernuzzi}\ \emph
  {et~al.}(2015{\natexlab{b}})\citenamefont {Bernuzzi}, \citenamefont {Nagar},
  \citenamefont {Dietrich},\ and\ \citenamefont {Damour}}]{Bernuzzi:2014owa}%
  \BibitemOpen
  \bibfield  {author} {\bibinfo {author} {\bibfnamefont {S.}~\bibnamefont
  {Bernuzzi}}, \bibinfo {author} {\bibfnamefont {A.}~\bibnamefont {Nagar}},
  \bibinfo {author} {\bibfnamefont {T.}~\bibnamefont {Dietrich}}, \ and\
  \bibinfo {author} {\bibfnamefont {T.}~\bibnamefont {Damour}},\ }\href
  {\doibase 10.1103/PhysRevLett.114.161103} {\bibfield  {journal} {\bibinfo
  {journal} {Phys.Rev.Lett.}\ }\textbf {\bibinfo {volume} {114}},\ \bibinfo
  {pages} {161103} (\bibinfo {year} {2015}{\natexlab{b}})},\ \Eprint
  {http://arxiv.org/abs/1412.4553} {arXiv:1412.4553 [gr-qc]} \BibitemShut
  {NoStop}%
\bibitem [{\citenamefont {Dietrich}\ \emph {et~al.}(2015)\citenamefont
  {Dietrich}, \citenamefont {Bernuzzi}, \citenamefont {Ujevic},\ and\
  \citenamefont {Br{\"u}gmann}}]{Dietrich:2015iva}%
  \BibitemOpen
  \bibfield  {author} {\bibinfo {author} {\bibfnamefont {T.}~\bibnamefont
  {Dietrich}}, \bibinfo {author} {\bibfnamefont {S.}~\bibnamefont {Bernuzzi}},
  \bibinfo {author} {\bibfnamefont {M.}~\bibnamefont {Ujevic}}, \ and\ \bibinfo
  {author} {\bibfnamefont {B.}~\bibnamefont {Br{\"u}gmann}},\ }\href {\doibase
  10.1103/PhysRevD.91.124041} {\bibfield  {journal} {\bibinfo  {journal} {Phys.
  Rev.}\ }\textbf {\bibinfo {volume} {D91}},\ \bibinfo {pages} {124041}
  (\bibinfo {year} {2015})},\ \Eprint {http://arxiv.org/abs/1504.01266}
  {arXiv:1504.01266 [gr-qc]} \BibitemShut {NoStop}%
\bibitem [{\citenamefont {Dietrich}\ and\ \citenamefont
  {Hinderer}(2017)}]{Dietrich:2017feu}%
  \BibitemOpen
  \bibfield  {author} {\bibinfo {author} {\bibfnamefont {T.}~\bibnamefont
  {Dietrich}}\ and\ \bibinfo {author} {\bibfnamefont {T.}~\bibnamefont
  {Hinderer}},\ }\href {\doibase 10.1103/PhysRevD.95.124006} {\bibfield
  {journal} {\bibinfo  {journal} {Phys. Rev.}\ }\textbf {\bibinfo {volume}
  {D95}},\ \bibinfo {pages} {124006} (\bibinfo {year} {2017})},\ \Eprint
  {http://arxiv.org/abs/1702.02053} {arXiv:1702.02053 [gr-qc]} \BibitemShut
  {NoStop}%
\bibitem [{\citenamefont {Dietrich}\ \emph
  {et~al.}(2017{\natexlab{a}})\citenamefont {Dietrich}, \citenamefont
  {Bernuzzi},\ and\ \citenamefont {Tichy}}]{Dietrich:2017aum}%
  \BibitemOpen
  \bibfield  {author} {\bibinfo {author} {\bibfnamefont {T.}~\bibnamefont
  {Dietrich}}, \bibinfo {author} {\bibfnamefont {S.}~\bibnamefont {Bernuzzi}},
  \ and\ \bibinfo {author} {\bibfnamefont {W.}~\bibnamefont {Tichy}},\
  }\href@noop {} {\  (\bibinfo {year} {2017}{\natexlab{a}})},\ \Eprint
  {http://arxiv.org/abs/1706.02969} {arXiv:1706.02969 [gr-qc]} \BibitemShut
  {NoStop}%
\bibitem [{\citenamefont {Radice}\ \emph
  {et~al.}(2017{\natexlab{a}})\citenamefont {Radice}, \citenamefont {Bernuzzi},
  \citenamefont {Del~Pozzo}, \citenamefont {Roberts},\ and\ \citenamefont
  {Ott}}]{Radice:2016rys}%
  \BibitemOpen
  \bibfield  {author} {\bibinfo {author} {\bibfnamefont {D.}~\bibnamefont
  {Radice}}, \bibinfo {author} {\bibfnamefont {S.}~\bibnamefont {Bernuzzi}},
  \bibinfo {author} {\bibfnamefont {W.}~\bibnamefont {Del~Pozzo}}, \bibinfo
  {author} {\bibfnamefont {L.~F.}\ \bibnamefont {Roberts}}, \ and\ \bibinfo
  {author} {\bibfnamefont {C.~D.}\ \bibnamefont {Ott}},\ }\href {\doibase
  10.3847/2041-8213/aa775f} {\bibfield  {journal} {\bibinfo  {journal}
  {Astrophys. J.}\ }\textbf {\bibinfo {volume} {842}},\ \bibinfo {pages} {L10}
  (\bibinfo {year} {2017}{\natexlab{a}})},\ \Eprint
  {http://arxiv.org/abs/1612.06429} {arXiv:1612.06429 [astro-ph.HE]}
  \BibitemShut {NoStop}%
\bibitem [{\citenamefont {Dietrich}\ \emph
  {et~al.}(2017{\natexlab{b}})\citenamefont {Dietrich}, \citenamefont {Ujevic},
  \citenamefont {Tichy}, \citenamefont {Bernuzzi},\ and\ \citenamefont
  {Br{\"u}gmann}}]{Dietrich:2016hky}%
  \BibitemOpen
  \bibfield  {author} {\bibinfo {author} {\bibfnamefont {T.}~\bibnamefont
  {Dietrich}}, \bibinfo {author} {\bibfnamefont {M.}~\bibnamefont {Ujevic}},
  \bibinfo {author} {\bibfnamefont {W.}~\bibnamefont {Tichy}}, \bibinfo
  {author} {\bibfnamefont {S.}~\bibnamefont {Bernuzzi}}, \ and\ \bibinfo
  {author} {\bibfnamefont {B.}~\bibnamefont {Br{\"u}gmann}},\ }\href {\doibase
  10.1103/PhysRevD.95.024029} {\bibfield  {journal} {\bibinfo  {journal} {Phys.
  Rev.}\ }\textbf {\bibinfo {volume} {D95}},\ \bibinfo {pages} {024029}
  (\bibinfo {year} {2017}{\natexlab{b}})},\ \Eprint
  {http://arxiv.org/abs/1607.06636} {arXiv:1607.06636 [gr-qc]} \BibitemShut
  {NoStop}%
\bibitem [{\citenamefont {Dietrich}\ \emph
  {et~al.}(2017{\natexlab{c}})\citenamefont {Dietrich}, \citenamefont
  {Bernuzzi}, \citenamefont {Ujevic},\ and\ \citenamefont
  {Tichy}}]{Dietrich:2016lyp}%
  \BibitemOpen
  \bibfield  {author} {\bibinfo {author} {\bibfnamefont {T.}~\bibnamefont
  {Dietrich}}, \bibinfo {author} {\bibfnamefont {S.}~\bibnamefont {Bernuzzi}},
  \bibinfo {author} {\bibfnamefont {M.}~\bibnamefont {Ujevic}}, \ and\ \bibinfo
  {author} {\bibfnamefont {W.}~\bibnamefont {Tichy}},\ }\href {\doibase
  10.1103/PhysRevD.95.044045} {\bibfield  {journal} {\bibinfo  {journal} {Phys.
  Rev.}\ }\textbf {\bibinfo {volume} {D95}},\ \bibinfo {pages} {044045}
  (\bibinfo {year} {2017}{\natexlab{c}})},\ \Eprint
  {http://arxiv.org/abs/1611.07367} {arXiv:1611.07367 [gr-qc]} \BibitemShut
  {NoStop}%
\bibitem [{\citenamefont {Radice}(2017)}]{Radice:2017zta}%
  \BibitemOpen
  \bibfield  {author} {\bibinfo {author} {\bibfnamefont {D.}~\bibnamefont
  {Radice}},\ }\href {\doibase 10.3847/2041-8213/aa6483} {\bibfield  {journal}
  {\bibinfo  {journal} {Astrophys. J.}\ }\textbf {\bibinfo {volume} {838}},\
  \bibinfo {pages} {L2} (\bibinfo {year} {2017})},\ \Eprint
  {http://arxiv.org/abs/1703.02046} {arXiv:1703.02046 [astro-ph.HE]}
  \BibitemShut {NoStop}%
\bibitem [{\citenamefont {Radice}\ \emph
  {et~al.}(2017{\natexlab{b}})\citenamefont {Radice}, \citenamefont {Perego},\
  and\ \citenamefont {Zappa}}]{Radice:2017lry}%
  \BibitemOpen
  \bibfield  {author} {\bibinfo {author} {\bibfnamefont {D.}~\bibnamefont
  {Radice}}, \bibinfo {author} {\bibfnamefont {A.}~\bibnamefont {Perego}}, \
  and\ \bibinfo {author} {\bibfnamefont {F.}~\bibnamefont {Zappa}},\
  }\href@noop {} {\  (\bibinfo {year} {2017}{\natexlab{b}})},\ \Eprint
  {http://arxiv.org/abs/1711.03647} {arXiv:1711.03647 [astro-ph.HE]}
  \BibitemShut {NoStop}%
\bibitem [{\citenamefont {Radice}\ \emph {et~al.}()\citenamefont {Radice},
  \citenamefont {Perego}, \citenamefont {Bernuzzi},\ and\ \citenamefont
  {Roberts}}]{RadiceInPrep}%
  \BibitemOpen
  \bibfield  {author} {\bibinfo {author} {\bibfnamefont {D.}~\bibnamefont
  {Radice}}, \bibinfo {author} {\bibfnamefont {A.}~\bibnamefont {Perego}},
  \bibinfo {author} {\bibfnamefont {S.}~\bibnamefont {Bernuzzi}}, \ and\
  \bibinfo {author} {\bibfnamefont {L.~F.}\ \bibnamefont {Roberts}},\
  }\href@noop {} {\bibinfo  {journal} {In Prep.}\ }\BibitemShut {NoStop}%
\bibitem [{\citenamefont {Abbott}\ \emph
  {et~al.}(2016{\natexlab{a}})\citenamefont {Abbott} \emph
  {et~al.}}]{Abbott:2016blz}%
  \BibitemOpen
\bibfield  {journal} {  }\bibfield  {author} {\bibinfo {author} {\bibfnamefont
  {B.~P.}\ \bibnamefont {Abbott}} \emph {et~al.} (\bibinfo {collaboration}
  {Virgo, LIGO Scientific}),\ }\href {\doibase 10.1103/PhysRevLett.116.061102}
  {\bibfield  {journal} {\bibinfo  {journal} {Phys. Rev. Lett.}\ }\textbf
  {\bibinfo {volume} {116}},\ \bibinfo {pages} {061102} (\bibinfo {year}
  {2016}{\natexlab{a}})},\ \Eprint {http://arxiv.org/abs/1602.03837}
  {arXiv:1602.03837 [gr-qc]} \BibitemShut {NoStop}%
\bibitem [{\citenamefont {Abbott}\ \emph
  {et~al.}(2016{\natexlab{b}})\citenamefont {Abbott} \emph
  {et~al.}}]{Abbott:2016nmj}%
  \BibitemOpen
  \bibfield  {author} {\bibinfo {author} {\bibfnamefont {B.~P.}\ \bibnamefont
  {Abbott}} \emph {et~al.} (\bibinfo {collaboration} {Virgo, LIGO
  Scientific}),\ }\href {\doibase 10.1103/PhysRevLett.116.241103} {\bibfield
  {journal} {\bibinfo  {journal} {Phys. Rev. Lett.}\ }\textbf {\bibinfo
  {volume} {116}},\ \bibinfo {pages} {241103} (\bibinfo {year}
  {2016}{\natexlab{b}})},\ \Eprint {http://arxiv.org/abs/1606.04855}
  {arXiv:1606.04855 [gr-qc]} \BibitemShut {NoStop}%
\bibitem [{\citenamefont {Abbott}\ \emph
  {et~al.}(2017{\natexlab{b}})\citenamefont {Abbott} \emph
  {et~al.}}]{Abbott:2017vtc}%
  \BibitemOpen
  \bibfield  {author} {\bibinfo {author} {\bibfnamefont {B.~P.}\ \bibnamefont
  {Abbott}} \emph {et~al.} (\bibinfo {collaboration} {VIRGO, LIGO
  Scientific}),\ }\href {\doibase 10.1103/PhysRevLett.118.221101} {\bibfield
  {journal} {\bibinfo  {journal} {Phys. Rev. Lett.}\ }\textbf {\bibinfo
  {volume} {118}},\ \bibinfo {pages} {221101} (\bibinfo {year}
  {2017}{\natexlab{b}})},\ \Eprint {http://arxiv.org/abs/1706.01812}
  {arXiv:1706.01812 [gr-qc]} \BibitemShut {NoStop}%
\bibitem [{\citenamefont {Baker}\ \emph {et~al.}(2008)\citenamefont {Baker}
  \emph {et~al.}}]{Baker:2008mj}%
  \BibitemOpen
  \bibfield  {author} {\bibinfo {author} {\bibfnamefont {J.~G.}\ \bibnamefont
  {Baker}} \emph {et~al.},\ }\href {\doibase 10.1103/PhysRevD.78.044046}
  {\bibfield  {journal} {\bibinfo  {journal} {Phys. Rev.}\ }\textbf {\bibinfo
  {volume} {D78}},\ \bibinfo {pages} {044046} (\bibinfo {year} {2008})},\
  \Eprint {http://arxiv.org/abs/0805.1428} {arXiv:0805.1428 [gr-qc]}
  \BibitemShut {NoStop}%
\bibitem [{\citenamefont {Keitel}\ \emph {et~al.}(2017)\citenamefont {Keitel}
  \emph {et~al.}}]{Keitel:2016krm}%
  \BibitemOpen
  \bibfield  {author} {\bibinfo {author} {\bibfnamefont {D.}~\bibnamefont
  {Keitel}} \emph {et~al.},\ }\href {\doibase 10.1103/PhysRevD.96.024006}
  {\bibfield  {journal} {\bibinfo  {journal} {Phys. Rev.}\ }\textbf {\bibinfo
  {volume} {D96}},\ \bibinfo {pages} {024006} (\bibinfo {year} {2017})},\
  \Eprint {http://arxiv.org/abs/1612.09566} {arXiv:1612.09566 [gr-qc]}
  \BibitemShut {NoStop}%
\bibitem [{\citenamefont {Jiménez-Forteza}\ \emph {et~al.}(2017)\citenamefont
  {Jiménez-Forteza}, \citenamefont {Keitel}, \citenamefont {Husa},
  \citenamefont {Hannam}, \citenamefont {Khan},\ and\ \citenamefont
  {Pürrer}}]{Jimenez-Forteza:2016oae}%
  \BibitemOpen
  \bibfield  {author} {\bibinfo {author} {\bibfnamefont {X.}~\bibnamefont
  {Jiménez-Forteza}}, \bibinfo {author} {\bibfnamefont {D.}~\bibnamefont
  {Keitel}}, \bibinfo {author} {\bibfnamefont {S.}~\bibnamefont {Husa}},
  \bibinfo {author} {\bibfnamefont {M.}~\bibnamefont {Hannam}}, \bibinfo
  {author} {\bibfnamefont {S.}~\bibnamefont {Khan}}, \ and\ \bibinfo {author}
  {\bibfnamefont {M.}~\bibnamefont {Pürrer}},\ }\href {\doibase
  10.1103/PhysRevD.95.064024} {\bibfield  {journal} {\bibinfo  {journal} {Phys.
  Rev.}\ }\textbf {\bibinfo {volume} {D95}},\ \bibinfo {pages} {064024}
  (\bibinfo {year} {2017})},\ \Eprint {http://arxiv.org/abs/1611.00332}
  {arXiv:1611.00332 [gr-qc]} \BibitemShut {NoStop}%
\bibitem [{\citenamefont {Healy}\ and\ \citenamefont
  {Lousto}(2017)}]{Healy:2016lce}%
  \BibitemOpen
  \bibfield  {author} {\bibinfo {author} {\bibfnamefont {J.}~\bibnamefont
  {Healy}}\ and\ \bibinfo {author} {\bibfnamefont {C.~O.}\ \bibnamefont
  {Lousto}},\ }\href {\doibase 10.1103/PhysRevD.95.024037} {\bibfield
  {journal} {\bibinfo  {journal} {Phys. Rev.}\ }\textbf {\bibinfo {volume}
  {D95}},\ \bibinfo {pages} {024037} (\bibinfo {year} {2017})},\ \Eprint
  {http://arxiv.org/abs/1610.09713} {arXiv:1610.09713 [gr-qc]} \BibitemShut
  {NoStop}%
\bibitem [{\citenamefont {Baiotti}\ \emph {et~al.}(2008)\citenamefont
  {Baiotti}, \citenamefont {Giacomazzo},\ and\ \citenamefont
  {Rezzolla}}]{Baiotti:2008ra}%
  \BibitemOpen
  \bibfield  {author} {\bibinfo {author} {\bibfnamefont {L.}~\bibnamefont
  {Baiotti}}, \bibinfo {author} {\bibfnamefont {B.}~\bibnamefont {Giacomazzo}},
  \ and\ \bibinfo {author} {\bibfnamefont {L.}~\bibnamefont {Rezzolla}},\
  }\href {\doibase 10.1103/PhysRevD.78.084033} {\bibfield  {journal} {\bibinfo
  {journal} {Phys. Rev.}\ }\textbf {\bibinfo {volume} {D78}},\ \bibinfo {pages}
  {084033} (\bibinfo {year} {2008})},\ \Eprint {http://arxiv.org/abs/0804.0594}
  {arXiv:0804.0594 [gr-qc]} \BibitemShut {NoStop}%
\bibitem [{\citenamefont {Tichy}(2011)}]{Tichy:2011gw}%
  \BibitemOpen
  \bibfield  {author} {\bibinfo {author} {\bibfnamefont {W.}~\bibnamefont
  {Tichy}},\ }\href {\doibase 10.1103/PhysRevD.84.024041} {\bibfield  {journal}
  {\bibinfo  {journal} {Phys.Rev.}\ }\textbf {\bibinfo {volume} {D84}},\
  \bibinfo {pages} {024041} (\bibinfo {year} {2011})},\ \Eprint
  {http://arxiv.org/abs/1107.1440} {arXiv:1107.1440 [gr-qc]} \BibitemShut
  {NoStop}%
\bibitem [{\citenamefont {Radice}\ \emph {et~al.}(2016)\citenamefont {Radice},
  \citenamefont {Galeazzi}, \citenamefont {Lippuner}, \citenamefont {Roberts},
  \citenamefont {Ott},\ and\ \citenamefont {Rezzolla}}]{Radice:2016dwd}%
  \BibitemOpen
  \bibfield  {author} {\bibinfo {author} {\bibfnamefont {D.}~\bibnamefont
  {Radice}}, \bibinfo {author} {\bibfnamefont {F.}~\bibnamefont {Galeazzi}},
  \bibinfo {author} {\bibfnamefont {J.}~\bibnamefont {Lippuner}}, \bibinfo
  {author} {\bibfnamefont {L.~F.}\ \bibnamefont {Roberts}}, \bibinfo {author}
  {\bibfnamefont {C.~D.}\ \bibnamefont {Ott}}, \ and\ \bibinfo {author}
  {\bibfnamefont {L.}~\bibnamefont {Rezzolla}},\ }\href {\doibase
  10.1093/mnras/stw1227} {\bibfield  {journal} {\bibinfo  {journal} {Mon. Not.
  Roy. Astron. Soc.}\ }\textbf {\bibinfo {volume} {460}},\ \bibinfo {pages}
  {3255} (\bibinfo {year} {2016})},\ \Eprint {http://arxiv.org/abs/1601.02426}
  {arXiv:1601.02426 [astro-ph.HE]} \BibitemShut {NoStop}%
\bibitem [{\citenamefont {Br{\"u}gmann}\ \emph {et~al.}(2008)\citenamefont
  {Br{\"u}gmann}, \citenamefont {Gonzalez}, \citenamefont {Hannam},
  \citenamefont {Husa}, \citenamefont {Sperhake} \emph
  {et~al.}}]{Brugmann:2008zz}%
  \BibitemOpen
  \bibfield  {author} {\bibinfo {author} {\bibfnamefont {B.}~\bibnamefont
  {Br{\"u}gmann}}, \bibinfo {author} {\bibfnamefont {J.~A.}\ \bibnamefont
  {Gonzalez}}, \bibinfo {author} {\bibfnamefont {M.}~\bibnamefont {Hannam}},
  \bibinfo {author} {\bibfnamefont {S.}~\bibnamefont {Husa}}, \bibinfo {author}
  {\bibfnamefont {U.}~\bibnamefont {Sperhake}},  \emph {et~al.},\ }\href
  {\doibase 10.1103/PhysRevD.77.024027} {\bibfield  {journal} {\bibinfo
  {journal} {Phys.Rev.}\ }\textbf {\bibinfo {volume} {D77}},\ \bibinfo {pages}
  {024027} (\bibinfo {year} {2008})},\ \Eprint
  {http://arxiv.org/abs/gr-qc/0610128} {arXiv:gr-qc/0610128 [gr-qc]}
  \BibitemShut {NoStop}%
\bibitem [{\citenamefont {Thierfelder}\ \emph {et~al.}(2011)\citenamefont
  {Thierfelder}, \citenamefont {Bernuzzi},\ and\ \citenamefont
  {Br{\"u}gmann}}]{Thierfelder:2011yi}%
  \BibitemOpen
  \bibfield  {author} {\bibinfo {author} {\bibfnamefont {M.}~\bibnamefont
  {Thierfelder}}, \bibinfo {author} {\bibfnamefont {S.}~\bibnamefont
  {Bernuzzi}}, \ and\ \bibinfo {author} {\bibfnamefont {B.}~\bibnamefont
  {Br{\"u}gmann}},\ }\href {\doibase 10.1103/PhysRevD.84.044012} {\bibfield
  {journal} {\bibinfo  {journal} {Phys.Rev.}\ }\textbf {\bibinfo {volume}
  {D84}},\ \bibinfo {pages} {044012} (\bibinfo {year} {2011})},\ \Eprint
  {http://arxiv.org/abs/1104.4751} {arXiv:1104.4751 [gr-qc]} \BibitemShut
  {NoStop}%
\bibitem [{\citenamefont {Radice}\ \emph
  {et~al.}(2014{\natexlab{a}})\citenamefont {Radice}, \citenamefont
  {Rezzolla},\ and\ \citenamefont {Galeazzi}}]{radice:2013hxh}%
  \BibitemOpen
  \bibfield  {author} {\bibinfo {author} {\bibfnamefont {D.}~\bibnamefont
  {Radice}}, \bibinfo {author} {\bibfnamefont {L.}~\bibnamefont {Rezzolla}}, \
  and\ \bibinfo {author} {\bibfnamefont {F.}~\bibnamefont {Galeazzi}},\ }\href
  {\doibase 10.1093/mnrasl/slt137} {\bibfield  {journal} {\bibinfo  {journal}
  {Mon.Not.Roy.Astron.Soc.}\ }\textbf {\bibinfo {volume} {437}},\ \bibinfo
  {pages} {L46} (\bibinfo {year} {2014}{\natexlab{a}})},\ \Eprint
  {http://arxiv.org/abs/1306.6052} {arXiv:1306.6052 [gr-qc]} \BibitemShut
  {NoStop}%
\bibitem [{\citenamefont {Radice}\ \emph
  {et~al.}(2014{\natexlab{b}})\citenamefont {Radice}, \citenamefont
  {Rezzolla},\ and\ \citenamefont {Galeazzi}}]{radice:2013xpa}%
  \BibitemOpen
  \bibfield  {author} {\bibinfo {author} {\bibfnamefont {D.}~\bibnamefont
  {Radice}}, \bibinfo {author} {\bibfnamefont {L.}~\bibnamefont {Rezzolla}}, \
  and\ \bibinfo {author} {\bibfnamefont {F.}~\bibnamefont {Galeazzi}},\ }\href
  {\doibase 10.1088/0264-9381/31/7/075012} {\bibfield  {journal} {\bibinfo
  {journal} {Class.Quant.Grav.}\ }\textbf {\bibinfo {volume} {31}},\ \bibinfo
  {pages} {075012} (\bibinfo {year} {2014}{\natexlab{b}})},\ \Eprint
  {http://arxiv.org/abs/1312.5004} {arXiv:1312.5004 [gr-qc]} \BibitemShut
  {NoStop}%
\bibitem [{\citenamefont {Dietrich}\ \emph {et~al.}()\citenamefont {Dietrich}
  \emph {et~al.}}]{DietrichInPrep}%
  \BibitemOpen
  \bibfield  {author} {\bibinfo {author} {\bibfnamefont {T.}~\bibnamefont
  {Dietrich}} \emph {et~al.},\ }\href@noop {} {\bibinfo  {journal} {In Prep.}\
  }\BibitemShut {NoStop}%
\bibitem [{\citenamefont {Damour}\ \emph
  {et~al.}(2012{\natexlab{a}})\citenamefont {Damour}, \citenamefont {Nagar},
  \citenamefont {Pollney},\ and\ \citenamefont {Reisswig}}]{Damour:2011fu}%
  \BibitemOpen
\bibfield  {journal} {  }\bibfield  {author} {\bibinfo {author} {\bibfnamefont
  {T.}~\bibnamefont {Damour}}, \bibinfo {author} {\bibfnamefont
  {A.}~\bibnamefont {Nagar}}, \bibinfo {author} {\bibfnamefont
  {D.}~\bibnamefont {Pollney}}, \ and\ \bibinfo {author} {\bibfnamefont
  {C.}~\bibnamefont {Reisswig}},\ }\href {\doibase
  10.1103/PhysRevLett.108.131101} {\bibfield  {journal} {\bibinfo  {journal}
  {Phys.Rev.Lett.}\ }\textbf {\bibinfo {volume} {108}},\ \bibinfo {pages}
  {131101} (\bibinfo {year} {2012}{\natexlab{a}})},\ \Eprint
  {http://arxiv.org/abs/1110.2938} {arXiv:1110.2938 [gr-qc]} \BibitemShut
  {NoStop}%
\bibitem [{\citenamefont {Damour}\ and\ \citenamefont
  {Nagar}(2010)}]{Damour:2009wj}%
  \BibitemOpen
  \bibfield  {author} {\bibinfo {author} {\bibfnamefont {T.}~\bibnamefont
  {Damour}}\ and\ \bibinfo {author} {\bibfnamefont {A.}~\bibnamefont {Nagar}},\
  }\href {\doibase 10.1103/PhysRevD.81.084016} {\bibfield  {journal} {\bibinfo
  {journal} {Phys. Rev.}\ }\textbf {\bibinfo {volume} {D81}},\ \bibinfo {pages}
  {084016} (\bibinfo {year} {2010})},\ \Eprint {http://arxiv.org/abs/0911.5041}
  {arXiv:0911.5041 [gr-qc]} \BibitemShut {NoStop}%
\bibitem [{\citenamefont {Damour}\ \emph
  {et~al.}(2012{\natexlab{b}})\citenamefont {Damour}, \citenamefont {Nagar},\
  and\ \citenamefont {Villain}}]{Damour:2012yf}%
  \BibitemOpen
  \bibfield  {author} {\bibinfo {author} {\bibfnamefont {T.}~\bibnamefont
  {Damour}}, \bibinfo {author} {\bibfnamefont {A.}~\bibnamefont {Nagar}}, \
  and\ \bibinfo {author} {\bibfnamefont {L.}~\bibnamefont {Villain}},\ }\href
  {\doibase 10.1103/PhysRevD.85.123007} {\bibfield  {journal} {\bibinfo
  {journal} {Phys.Rev.}\ }\textbf {\bibinfo {volume} {D85}},\ \bibinfo {pages}
  {123007} (\bibinfo {year} {2012}{\natexlab{b}})},\ \Eprint
  {http://arxiv.org/abs/1203.4352} {arXiv:1203.4352 [gr-qc]} \BibitemShut
  {NoStop}%
\bibitem [{\citenamefont {Baumgarte}\ \emph {et~al.}(2000)\citenamefont
  {Baumgarte}, \citenamefont {Shapiro},\ and\ \citenamefont
  {Shibata}}]{Baumgarte:1999cq}%
  \BibitemOpen
  \bibfield  {author} {\bibinfo {author} {\bibfnamefont {T.~W.}\ \bibnamefont
  {Baumgarte}}, \bibinfo {author} {\bibfnamefont {S.~L.}\ \bibnamefont
  {Shapiro}}, \ and\ \bibinfo {author} {\bibfnamefont {M.}~\bibnamefont
  {Shibata}},\ }\href@noop {} {\bibfield  {journal} {\bibinfo  {journal}
  {Astrophys. J.}\ }\textbf {\bibinfo {volume} {528}},\ \bibinfo {pages} {L29}
  (\bibinfo {year} {2000})},\ \Eprint {http://arxiv.org/abs/astro-ph/9910565}
  {arXiv:astro-ph/9910565} \BibitemShut {NoStop}%
\bibitem [{\citenamefont {Hotokezaka}\ \emph {et~al.}(2011)\citenamefont
  {Hotokezaka}, \citenamefont {Kyutoku}, \citenamefont {Okawa}, \citenamefont
  {Shibata},\ and\ \citenamefont {Kiuchi}}]{Hotokezaka:2011dh}%
  \BibitemOpen
  \bibfield  {author} {\bibinfo {author} {\bibfnamefont {K.}~\bibnamefont
  {Hotokezaka}}, \bibinfo {author} {\bibfnamefont {K.}~\bibnamefont {Kyutoku}},
  \bibinfo {author} {\bibfnamefont {H.}~\bibnamefont {Okawa}}, \bibinfo
  {author} {\bibfnamefont {M.}~\bibnamefont {Shibata}}, \ and\ \bibinfo
  {author} {\bibfnamefont {K.}~\bibnamefont {Kiuchi}},\ }\href {\doibase
  10.1103/PhysRevD.83.124008} {\bibfield  {journal} {\bibinfo  {journal}
  {Phys.Rev.}\ }\textbf {\bibinfo {volume} {D83}},\ \bibinfo {pages} {124008}
  (\bibinfo {year} {2011})},\ \Eprint {http://arxiv.org/abs/1105.4370}
  {arXiv:1105.4370 [astro-ph.HE]} \BibitemShut {NoStop}%
\bibitem [{\citenamefont {Bauswein}\ \emph {et~al.}(2013)\citenamefont
  {Bauswein}, \citenamefont {Baumgarte},\ and\ \citenamefont
  {Janka}}]{Bauswein:2013jpa}%
  \BibitemOpen
  \bibfield  {author} {\bibinfo {author} {\bibfnamefont {A.}~\bibnamefont
  {Bauswein}}, \bibinfo {author} {\bibfnamefont {T.}~\bibnamefont {Baumgarte}},
  \ and\ \bibinfo {author} {\bibfnamefont {H.~T.}\ \bibnamefont {Janka}},\
  }\href {\doibase 10.1103/PhysRevLett.111.131101} {\bibfield  {journal}
  {\bibinfo  {journal} {Phys.Rev.Lett.}\ }\textbf {\bibinfo {volume} {111}},\
  \bibinfo {pages} {131101} (\bibinfo {year} {2013})},\ \Eprint
  {http://arxiv.org/abs/1307.5191} {arXiv:1307.5191 [astro-ph.SR]} \BibitemShut
  {NoStop}%
\bibitem [{\citenamefont {Hinderer}\ \emph {et~al.}(2010)\citenamefont
  {Hinderer}, \citenamefont {Lackey}, \citenamefont {Lang},\ and\ \citenamefont
  {Read}}]{Hinderer:2009ca}%
  \BibitemOpen
  \bibfield  {author} {\bibinfo {author} {\bibfnamefont {T.}~\bibnamefont
  {Hinderer}}, \bibinfo {author} {\bibfnamefont {B.~D.}\ \bibnamefont
  {Lackey}}, \bibinfo {author} {\bibfnamefont {R.~N.}\ \bibnamefont {Lang}}, \
  and\ \bibinfo {author} {\bibfnamefont {J.~S.}\ \bibnamefont {Read}},\ }\href
  {\doibase 10.1103/PhysRevD.81.123016} {\bibfield  {journal} {\bibinfo
  {journal} {Phys. Rev.}\ }\textbf {\bibinfo {volume} {D81}},\ \bibinfo {pages}
  {123016} (\bibinfo {year} {2010})},\ \Eprint {http://arxiv.org/abs/0911.3535}
  {arXiv:0911.3535 [astro-ph.HE]} \BibitemShut {NoStop}%
\bibitem [{\citenamefont {Hotokezaka}\ \emph
  {et~al.}(2013{\natexlab{a}})\citenamefont {Hotokezaka}, \citenamefont
  {Kiuchi}, \citenamefont {Kyutoku}, \citenamefont {Okawa}, \citenamefont
  {Sekiguchi} \emph {et~al.}}]{Hotokezaka:2012ze}%
  \BibitemOpen
  \bibfield  {author} {\bibinfo {author} {\bibfnamefont {K.}~\bibnamefont
  {Hotokezaka}}, \bibinfo {author} {\bibfnamefont {K.}~\bibnamefont {Kiuchi}},
  \bibinfo {author} {\bibfnamefont {K.}~\bibnamefont {Kyutoku}}, \bibinfo
  {author} {\bibfnamefont {H.}~\bibnamefont {Okawa}}, \bibinfo {author}
  {\bibfnamefont {Y.-i.}\ \bibnamefont {Sekiguchi}},  \emph {et~al.},\ }\href
  {\doibase 10.1103/PhysRevD.87.024001} {\bibfield  {journal} {\bibinfo
  {journal} {Phys.Rev.}\ }\textbf {\bibinfo {volume} {D87}},\ \bibinfo {pages}
  {024001} (\bibinfo {year} {2013}{\natexlab{a}})},\ \Eprint
  {http://arxiv.org/abs/1212.0905} {arXiv:1212.0905 [astro-ph.HE]} \BibitemShut
  {NoStop}%
\bibitem [{\citenamefont {Hotokezaka}\ \emph
  {et~al.}(2013{\natexlab{b}})\citenamefont {Hotokezaka}, \citenamefont
  {Kiuchi}, \citenamefont {Kyutoku}, \citenamefont {Muranushi}, \citenamefont
  {Sekiguchi} \emph {et~al.}}]{Hotokezaka:2013iia}%
  \BibitemOpen
  \bibfield  {author} {\bibinfo {author} {\bibfnamefont {K.}~\bibnamefont
  {Hotokezaka}}, \bibinfo {author} {\bibfnamefont {K.}~\bibnamefont {Kiuchi}},
  \bibinfo {author} {\bibfnamefont {K.}~\bibnamefont {Kyutoku}}, \bibinfo
  {author} {\bibfnamefont {T.}~\bibnamefont {Muranushi}}, \bibinfo {author}
  {\bibfnamefont {Y.-i.}\ \bibnamefont {Sekiguchi}},  \emph {et~al.},\ }\href
  {\doibase 10.1103/PhysRevD.88.044026} {\bibfield  {journal} {\bibinfo
  {journal} {Phys.Rev.}\ }\textbf {\bibinfo {volume} {D88}},\ \bibinfo {pages}
  {044026} (\bibinfo {year} {2013}{\natexlab{b}})},\ \Eprint
  {http://arxiv.org/abs/1307.5888} {arXiv:1307.5888 [astro-ph.HE]} \BibitemShut
  {NoStop}%
\bibitem [{\citenamefont {Bernuzzi}\ \emph {et~al.}(2016)\citenamefont
  {Bernuzzi}, \citenamefont {Radice}, \citenamefont {Ott}, \citenamefont
  {Roberts}, \citenamefont {Moesta},\ and\ \citenamefont
  {Galeazzi}}]{Bernuzzi:2015opx}%
  \BibitemOpen
  \bibfield  {author} {\bibinfo {author} {\bibfnamefont {S.}~\bibnamefont
  {Bernuzzi}}, \bibinfo {author} {\bibfnamefont {D.}~\bibnamefont {Radice}},
  \bibinfo {author} {\bibfnamefont {C.~D.}\ \bibnamefont {Ott}}, \bibinfo
  {author} {\bibfnamefont {L.~F.}\ \bibnamefont {Roberts}}, \bibinfo {author}
  {\bibfnamefont {P.}~\bibnamefont {Moesta}}, \ and\ \bibinfo {author}
  {\bibfnamefont {F.}~\bibnamefont {Galeazzi}},\ }\href {\doibase
  10.1103/PhysRevD.94.024023} {\bibfield  {journal} {\bibinfo  {journal} {Phys.
  Rev.}\ }\textbf {\bibinfo {volume} {D94}},\ \bibinfo {pages} {024023}
  (\bibinfo {year} {2016})},\ \Eprint {http://arxiv.org/abs/1512.06397}
  {arXiv:1512.06397 [gr-qc]} \BibitemShut {NoStop}%
\bibitem [{\citenamefont {Abbott}\ \emph
  {et~al.}(2017{\natexlab{c}})\citenamefont {Abbott} \emph
  {et~al.}}]{Abbott:2017eaw}%
  \BibitemOpen
  \bibfield  {author} {\bibinfo {author} {\bibfnamefont {B.~P.}\ \bibnamefont
  {Abbott}} \emph {et~al.} (\bibinfo {collaboration} {Virgo, LIGO
  Scientific}),\ }\href@noop {} {\  (\bibinfo {year} {2017}{\natexlab{c}})},\
  \Eprint {http://arxiv.org/abs/1710.09320} {arXiv:1710.09320 [astro-ph.HE]}
  \BibitemShut {NoStop}%
\bibitem [{\citenamefont {Shibata}\ and\ \citenamefont
  {Taniguchi}(2006)}]{Shibata:2006nm}%
  \BibitemOpen
  \bibfield  {author} {\bibinfo {author} {\bibfnamefont {M.}~\bibnamefont
  {Shibata}}\ and\ \bibinfo {author} {\bibfnamefont {K.}~\bibnamefont
  {Taniguchi}},\ }\href {\doibase 10.1103/PhysRevD.73.064027} {\bibfield
  {journal} {\bibinfo  {journal} {Phys.Rev.}\ }\textbf {\bibinfo {volume}
  {D73}},\ \bibinfo {pages} {064027} (\bibinfo {year} {2006})},\ \Eprint
  {http://arxiv.org/abs/astro-ph/0603145} {arXiv:astro-ph/0603145 [astro-ph]}
  \BibitemShut {NoStop}%
\bibitem [{\citenamefont {Rezzolla}\ \emph {et~al.}(2010)\citenamefont
  {Rezzolla}, \citenamefont {Baiotti}, \citenamefont {Giacomazzo},
  \citenamefont {Link},\ and\ \citenamefont {Font}}]{Rezzolla:2010fd}%
  \BibitemOpen
  \bibfield  {author} {\bibinfo {author} {\bibfnamefont {L.}~\bibnamefont
  {Rezzolla}}, \bibinfo {author} {\bibfnamefont {L.}~\bibnamefont {Baiotti}},
  \bibinfo {author} {\bibfnamefont {B.}~\bibnamefont {Giacomazzo}}, \bibinfo
  {author} {\bibfnamefont {D.}~\bibnamefont {Link}}, \ and\ \bibinfo {author}
  {\bibfnamefont {J.~A.}\ \bibnamefont {Font}},\ }\href {\doibase
  10.1088/0264-9381/27/11/114105} {\bibfield  {journal} {\bibinfo  {journal}
  {Class. Quant. Grav.}\ }\textbf {\bibinfo {volume} {27}},\ \bibinfo {pages}
  {114105} (\bibinfo {year} {2010})},\ \Eprint {http://arxiv.org/abs/1001.3074}
  {arXiv:1001.3074 [gr-qc]} \BibitemShut {NoStop}%
\bibitem [{\citenamefont {Shibata}\ \emph {et~al.}(2017)\citenamefont
  {Shibata}, \citenamefont {Fujibayashi}, \citenamefont {Hotokezaka},
  \citenamefont {Kiuchi}, \citenamefont {Kyutoku}, \citenamefont {Sekiguchi},\
  and\ \citenamefont {Tanaka}}]{Shibata:2017xdx}%
  \BibitemOpen
  \bibfield  {author} {\bibinfo {author} {\bibfnamefont {M.}~\bibnamefont
  {Shibata}}, \bibinfo {author} {\bibfnamefont {S.}~\bibnamefont
  {Fujibayashi}}, \bibinfo {author} {\bibfnamefont {K.}~\bibnamefont
  {Hotokezaka}}, \bibinfo {author} {\bibfnamefont {K.}~\bibnamefont {Kiuchi}},
  \bibinfo {author} {\bibfnamefont {K.}~\bibnamefont {Kyutoku}}, \bibinfo
  {author} {\bibfnamefont {Y.}~\bibnamefont {Sekiguchi}}, \ and\ \bibinfo
  {author} {\bibfnamefont {M.}~\bibnamefont {Tanaka}},\ }\href@noop {} {\
  (\bibinfo {year} {2017})},\ \Eprint {http://arxiv.org/abs/1710.07579}
  {arXiv:1710.07579 [astro-ph.HE]} \BibitemShut {NoStop}%
\bibitem [{\citenamefont {Fujibayashi}\ \emph {et~al.}(2017)\citenamefont
  {Fujibayashi}, \citenamefont {Kiuchi}, \citenamefont {Nishimura},
  \citenamefont {Sekiguchi},\ and\ \citenamefont
  {Shibata}}]{Fujibayashi:2017puw}%
  \BibitemOpen
  \bibfield  {author} {\bibinfo {author} {\bibfnamefont {S.}~\bibnamefont
  {Fujibayashi}}, \bibinfo {author} {\bibfnamefont {K.}~\bibnamefont {Kiuchi}},
  \bibinfo {author} {\bibfnamefont {N.}~\bibnamefont {Nishimura}}, \bibinfo
  {author} {\bibfnamefont {Y.}~\bibnamefont {Sekiguchi}}, \ and\ \bibinfo
  {author} {\bibfnamefont {M.}~\bibnamefont {Shibata}},\ }\href@noop {} {\
  (\bibinfo {year} {2017})},\ \Eprint {http://arxiv.org/abs/1711.02093}
  {arXiv:1711.02093 [astro-ph.HE]} \BibitemShut {NoStop}%
\end{thebibliography}%

\end{document}